\documentclass[twoside]{article}
\usepackage{fleqn,espcrc2,epsfig}
\newcommand{\PS}{{\rm P}}
\newcommand{\VEC}{{\rm V}}
\newcommand{\be}{\begin{equation}}
\newcommand{\ee}{\end{equation}}
\usepackage{graphicx}
\usepackage[figuresright]{rotating}

\title{
\hfill\begin{minipage}{0pt}\scriptsize \begin{tabbing}
\hspace*{\fill} Edinburgh-2001/15\\ \end{tabbing}\end{minipage}\\[8pt]
Semileptonic decay of a heavy-light pseudoscalar to a light vector meson}
\author{UKQCD Collaboration\\
	Presented by James Gill\address{Department of Physics and Astronomy, University of Edinburgh EH9 3JZ, Scotland}} 
\begin{document}
\begin{abstract}
We present the results of a calculation of semileptonic form factors of $D$ and $B$ mesons.  The calculation uses nonperturbatively ${\cal O}(a)$ improved quenched lattice QCD and two values of the coupling.  Results for charm mesons show reasonable agreement with experiment.  The lattice results for $B\rightarrow\rho\ell\nu_\ell$ are compared with high $q^2$ experimental results.
\end{abstract}
\maketitle

\section{INTRODUCTION}
Semileptonic decays of heavy-light mesons are of interest because they can be used to determine elements of the CKM matrix.  The nonperturbative QCD information needed to understand a decay can be calculated in lattice QCD.  In the case of the decays $D\rightarrow K^*$ and $D_s\rightarrow\phi$ the branching ratios have been accurately measured and the relevant CKM matrix element ($V_{cs}$) is well determined by other methods \cite{pdg}.  These decays are a good test of lattice results.  The decay $B\rightarrow\rho$ depends on $|V_{ub}|$, which is poorly known, so lattice results for this decay are especially useful.

The weak matrix elements for the decay Pseudoscalar$\rightarrow$Vector ($P\rightarrow V$) can be parameterized in terms of 4 form factors.  Treating the $V^\mu$ and $A^\mu$ currents separately,
\begin{eqnarray}
\label{eq:mea}
\langle \VEC ,k,\eta|&&\hspace{-0.7cm}A_{\mu}|\PS,p
\rangle= i(m_\PS+m_\VEC)A_1(q^2) g_{\mu\nu} 
\eta^{\nu} \nonumber \\
&& -\frac{iA_2(q^2)}{m_\PS+m_\VEC} (p+k)_{\mu} 
q_\nu \eta^{\nu}\nonumber \\
&&+\frac{2im_\VEC A(q^2)}{q^2}(p-k)_{\mu}(p+k)_\nu \eta^{\nu}\\
\label{eq:mev}
\langle \VEC ,k,\eta|&&\hspace{-0.7cm}V_{\mu}|\PS,p\rangle= 
\frac{2 V(q^2)}{m_\PS+m_V} \varepsilon_{\mu\rho\sigma\nu}p^{\rho}k^{\sigma} \eta^{\nu}
\end{eqnarray}
where $k$ is the momentum of the $\VEC$, $p$ is the momentum of the $\PS$, $q=p-k$ and $\eta^\nu$ is the polarization vector.  Results are presented here for $A_0$ instead of $A$, where
\be
A_0=A-\frac{m_\PS+m_\VEC}{2m_\VEC}A_1+\frac{m_\PS-m_\VEC}{2m_\VEC}A_2 \ .
\ee
\section{SIMULATION DETAILS}
This calculation used the nonperturbatively ${\cal O}(a)$ improved action and the quenched approximation.  The coefficients used to improve and renormalize the currents were determined nonperturbatively in \cite{bhattacharya}.  Two values of the coupling were used, $\beta=6.0$ and $\beta=6.2$.  The lattice sizes were $16^3\times32$ and $24^3\times48$ respectively.

The calculation used the method described in \cite{bowler95}.  The 3 (or 1) form factors were obtained as free parameters in fits to three-point functions.  The fits were repeated with different $\vec p,\ \vec k$ combinations, which correspond to different values of $q^2$.  All cases with $|\vec p|,\ |\vec k|= 0\ \mathrm{or}\ 1$ in lattice units were considered.  However in some cases the three-point function was very noisy or did not seem to satisfy the assumption of the operators being well separated in time.  One momentum combination was rejected at $\beta=6.2$, several were rejected at $\beta=6.0$.

\section{CHARMED MESONS}
The simulation used 3 light quark masses around strange, and 4 heavy quark masses around charm.  The simulation form factors were extrapolated and interpolated to physical values of quark mass.  The values of the hopping parameter corresponding to massless, up/down and strange quarks were determined in \cite{ukqcd} (isospin symmetry was assumed).  The quark mass was improved using the boosted perturbation theory value of $b_m$.

The form factors were extrapolated in light quark mass using the ansatz
\be
\label{eq:ansatz1}
F(q^2, m_1,m_2)=x_0 + x_1 m_1 + x_2 m_2
\ee
where $F$ is a generic form factor, the $m_i$ are light quark masses, and the $x_i$ are free parameters.  The $x_i$ were determined separately for each form factor, momentum combination and heavy quark.  The ansatz (\ref{eq:ansatz1}) is a first-order Taylor expansion in $m_1,\ m_2$.  To a good approximation the light vector meson mass depends linearly on quark mass,  so (\ref{eq:ansatz1}) implicitly includes a first-order Taylor expansion in $m_\VEC$.

The form factors were interpolated in heavy quark mass to charm.  Different sensible choices of interpolation ansatz give almost exactly the same result,  because the heavy quark masses simulated are closely spaced around charm.  The results were interpolated in $m_\PS$ to the experimentally determined $D$ or $D_s$ mass.  This requires a determination of the lattice spacing, $a$.  $q^2$ at physical quark masses was calculated using the experimental values of meson mass and $a$.

\begin{figure}[ht!]
\vspace{-0.5cm}
\begin{center}
\epsfig{file=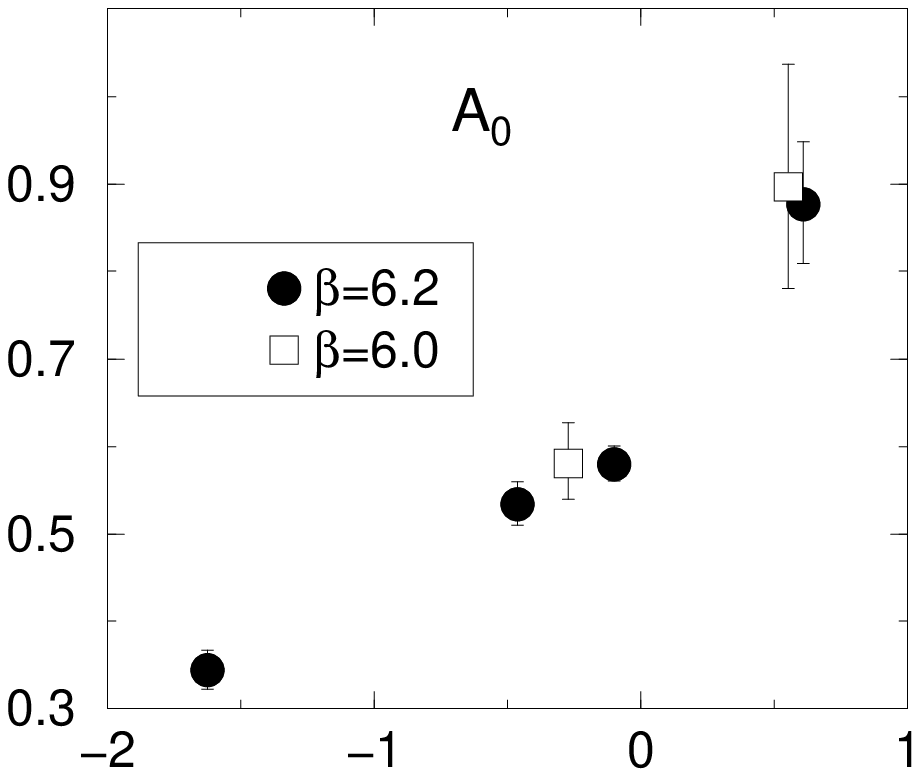,height=3.5cm,width=3.5cm}
\epsfig{file=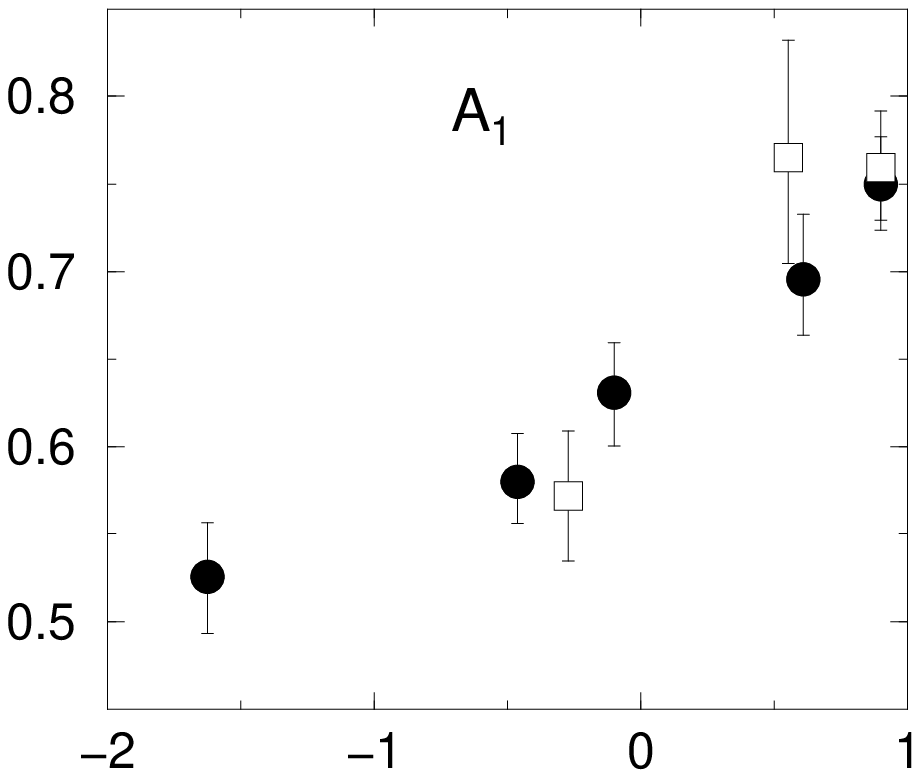,height=3.5cm,width=3.5cm}
\epsfig{file=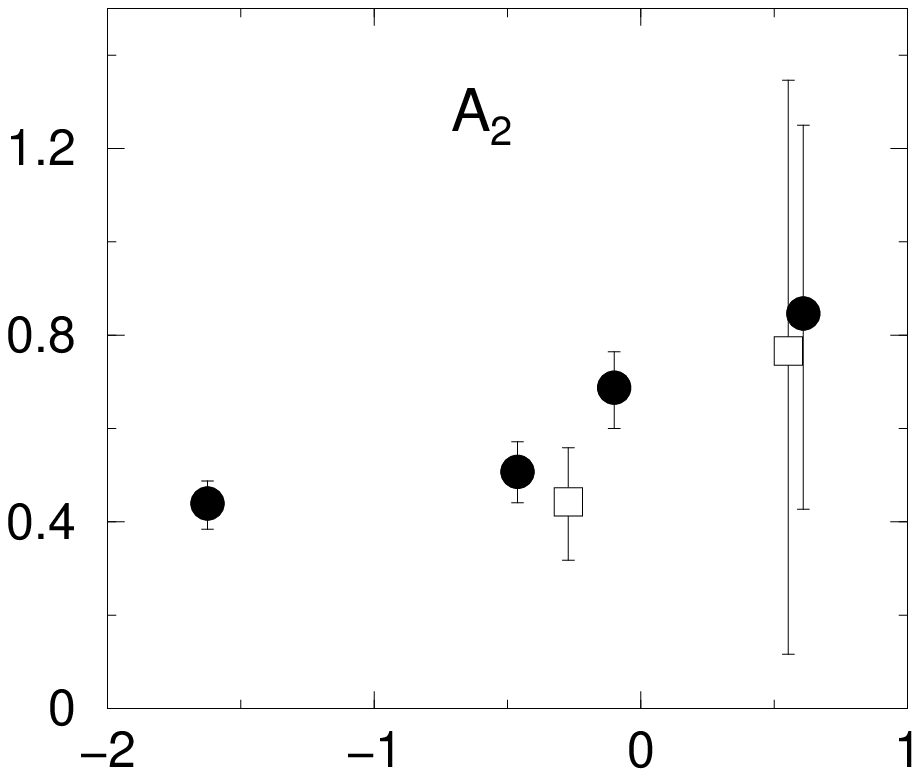,height=3.5cm,width=3.5cm}
\epsfig{file=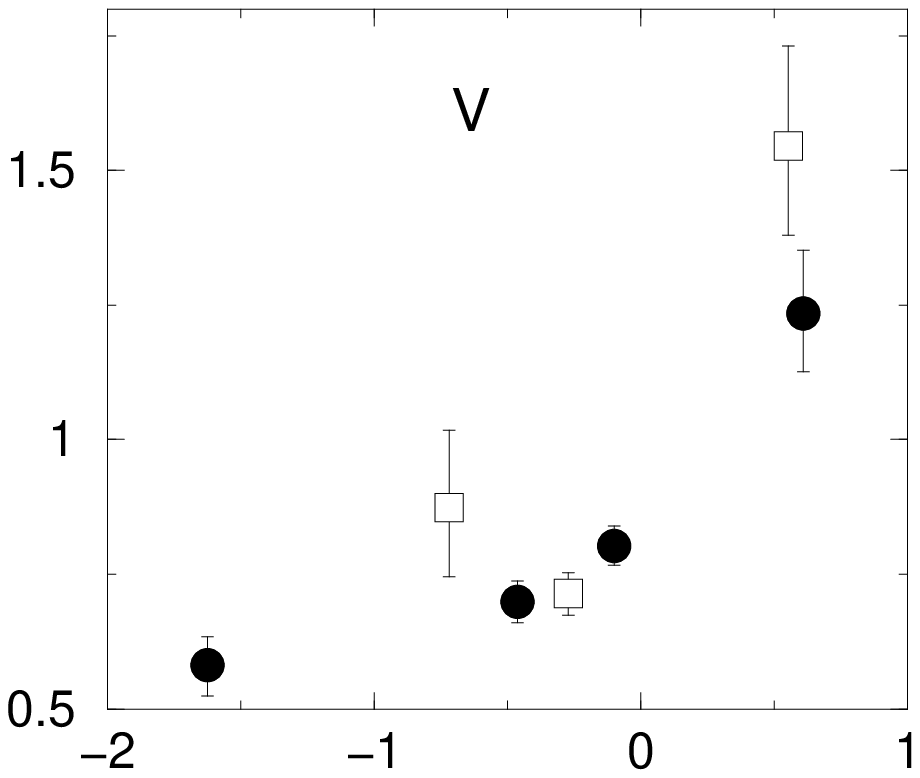,height=3.5cm,width=3.5cm}
$q^2\ (\mathrm{GeV}^2)$
\vspace{-0.7cm}
\caption{\label{fig:formfac}$D_s\rightarrow\phi$ form factors.}
\end{center}
\end{figure}

The statistical errors are smallest in the case $D_s\rightarrow\phi$ because this only involves interpolation of the data.  This is a good place to see if the results depend on $\beta$.  The results for $D_s\rightarrow\phi$ are shown in figure \ref{fig:formfac}.  The $\beta=6.0$ and $\beta=6.2$ results agree within errors, although unfortunately the $\beta=6.0$ results are rather noisy.  

With form factors at just two values of $\beta$ a reliable continuum extrapolation is not possible.  To compare lattice with experiment the $\beta=6.2$ results were treated as continuum results and the $\beta=6.0$ where discarded.  The form factors were interpolated in $q^2$ using a simple pole fit,
\be
F(q^2)=\frac{F(0)}{1-q^2/M_F^2} \ .
\ee
The axial form factors have the kinematic constraint $A(q^2)=0$, which was enforced in the fits.  The results for $A_2$ are too noisy to constrain a pole mass so this was assumed to be the same pole mass as for $A_1$ as suggested by pole dominance models \cite{stech}.  The high $q^2$ part of $A_2$ has very little influence on decay rate, so the lattice prediction for decay rate is not biased by this assumption.

\vspace{-0.8cm}

The results of the fits to the lattice form factors are shown in table \ref{tab:poles}.  Note that the results presented in this paper are preliminary.  The results shown use $r_0$ to set the scale, which gives $a^{-1}=2.913\ \mathrm{GeV}$.  To estimate systematic errors results were recalculated using $m_\rho$ to set the scale $(a^{-1}=2.54\ \mathrm{GeV})$.  The change in $F(0)$ was smaller than the statistical errors, in all cases.  Typically the $M_F^2$ are 10\% smaller, in physical units when $m_\rho$ is used to set the scale.  The lattice predictions for integrated decay rate are compared with experiment in table \ref{tab:intdecay}.  The leptons were assumed to be massless and the Zweig rule suppressed disconnected diagrams which contribute to the decay $D_s\rightarrow\phi$ were not included.  There is reasonable agreement between experiment and the lattice results.  The form factor ratios $V(0)/A_1(0)$ and $A_2(0)/A_1(0)$ have been measured for the decays $D_s\rightarrow\phi$ and $D\rightarrow K^*$.  In some cases there is a significant difference between the lattice and experimental ratio.
\begin{table}[ht!]
%\vspace{-0.6cm}
\begin{center}
\caption{\label{tab:poles} Parameters from pole fits to the lattice form factors.  Pole mass squared is written $M^2_F$ where $F$ is the form factor modelled by the pole.  Masses are in lattice units. Errors here and in table \ref{tab:intdecay} are statistical (68\% confidence level).}
\begin{tabular}{l c c c}\hline
&$D_s\rightarrow\phi$ &$D\rightarrow K^*$ &$D\rightarrow\rho$ \\ \hline 
$V(0)$      & $0.85^{+4}_{-3}$   & $0.80^{+5}_{-5}$  & $0.71^{+5}_{-6}$\\
$A_0(0)$    & $0.63^{+2}_{-2}$   & $0.64^{+3}_{-3}$  & $0.59^{+3}_{-3}$\\
$A_1(0)$    & $0.63^{+2}_{-2}$   & $0.65^{+2}_{-3}$  & $0.60^{+3}_{-3}$\\
$A_2(0)$    & $0.62^{+5}_{-5}$   & $0.67^{+7}_{-7}$  & $0.61^{+6}_{-7}$\\
$M^2_{V}$   & $0.53^{+5}_{-4}$   & $0.58^{+14}_{-8}$ & $0.57^{+14}_{-7}$\\
$M^2_{A_0}$ & $0.24^{+2}_{-2}$   & $0.28^{+5}_{-4}$  & $0.31^{+5}_{-4}$\\
$M^2_{A_1}$ & $0.69^{+13}_{-10}$ & $0.9^{+4}_{-2}$   & $1.0^{+6}_{-3}$ \\ \hline
\end{tabular}
%\vspace{-1cm}
\end{center}
\end{table}
\begin{table}[ht!]
\vspace{-0.6cm}
\begin{center}
\caption{\label{tab:intdecay}Integrated decay rates using the results in table \ref{tab:poles} and the PDG values for the relevant CKM matrix elements.  Units are $10^{10}s^{-1}$.}
\begin{tabular}{l c c}\hline
& Experiment & Lattice \\ \hline 
$D^+_s\rightarrow\phi \ell^+\nu_\ell$&$4.0\pm0.5$&$5.0\pm0.3$\\
$D^+\rightarrow \bar{K}^*(892)^0\ell^+\nu_\ell$&$4.5\pm0.4$&$5.5\pm0.5$\\
$D^+\rightarrow\rho^0e^+\nu_e$ &$0.21\pm0.08$&$0.19\pm0.02$\\\hline
%$D^+\rightarrow\rho^0\mu^+\nu_\mu$ & $0.26\pm0.07$&$0.19\pm0.02$\\\hline
\end{tabular}
\end{center}
\vspace{-0.5cm}
\end{table}

\section{EXTRAPOLATION TO $B\rightarrow\rho$}
To apply these lattice results to the decay $B\rightarrow\rho$ requires an extrapolation in heavy quark mass, after the light quark extrapolations.  The extrapolation is done at fixed $\omega$ where
\be
\omega=\frac{m_\PS^2+m_\VEC^2-q^2}{2m_\PS m_\VEC} .
\ee
The ansatz used is \cite{brho95}
\be
\label{hqet_extrap}
F(\omega,m_\PS)=m_\PS^N[\alpha_s(m_\PS)]^{-\frac{2}{11}} \Big(x_0+\frac{x_1}{m_\PS}+\frac{x_2}{m_\PS^2}\Big)
\ee
where $N=-\frac{1}{2}$ for $A_1$, $N=\frac{1}{2}$ otherwise.  According to heavy quark effective theory (\ref{hqet_extrap}) has the correct infinite quark mass limit for $\omega$ fixed and close to 1.  The $x_1$ and $x_2$ terms in (\ref{hqet_extrap}) are corrections due to the finite mass of the heavy quark.  Some of the simulated momentum combinations have a slightly different $\omega$ for the different heavy quark masses.  In these cases the form factors were interpolated in $q^2$ to give constant $\omega$.  In all cases the interpolation was a small shift in $q^2$ and form factor value.

After the extrapolation  $B\rightarrow\rho$ form factors were obtained at $q^2$ between about $14\ \mathrm{GeV}^2$ and $q^2_{max}$.  The results can be extrapolated in $q^2$ to obtain form factors for the whole kinematically allowed range and a prediction of integrated decay rate.  However the extrapolation results in a strongly model dependent result.  Instead we interpolate the lattice differential decay rate using
\be
\label{eq:dgdq}
\frac{1}{|V_{ub}|^2}\frac{d\Gamma}{dq^2}=
\frac{G_F^2q^2[\lambda(q^2)]^{\frac{1}{2}}}{192\pi^3m_B^3}
\Big(a+b(q^2-q^2_{max})\Big)
\ee
where $a,b$ are free parameters and $\lambda(q^2)$ is given in \cite{brho95}.  Note that (\ref{eq:dgdq}) is an approximtion and is not expected to be reliable much below $q^2=14\ \mathrm{GeV}^2$.  The fit to the lattice results is good, and  gives $a=38^{+8}_{-5}\ \mathrm{GeV}^2,\ \ b=0^{+2}_{-2}$ (statistical errors).  To estimate systematic errors the heavy quark extrapolation was repeated using the three heaviest masses and $x_2$ in (\ref{hqet_extrap}) fixed to 0.  This gives $a=33^{+7}_{-4}\ \mathrm{GeV}^2,\ \ b=1^{+1}_{-1}$.  The scale dependence of $a,b$ is insignificant.

Using $|V_{ub}|=3.5\times10^{-3}$ and the first set of values for $a,b$ gives the partially integrated decay rate $\Delta\Gamma(14\ \mathrm{GeV}^2<q^2<q^2_{max})=8.7^{+1.9}_{-1.2}\times10^{7}s^{-1}$.  This agrees with the experimental result $\Delta\Gamma(14\ \mathrm{GeV}^2<q^2<21\ \mathrm{GeV}^2)=7.1\pm2.4\times10^{7}s^{-1}$ \cite{cleo}.

This work was supported by PPARC and the European Community's Human potential programme under HPRN-CT-2000-00145 Hadrons/LatticeQCD.

\small
\begin{thebibliography}{10}
\bibitem{pdg}{D.~E.~Groom {\it et al.}}, Eur.~Phys.~J.~C, {\bf 15}, 1 (2000)
\bibitem{bhattacharya}{T.~Bhattacharya, R.~Gupta, W.~Lee, S.~Sharpe}, Phys.~Rev.~D {\bf 63}, 074505 (2001)
\bibitem{bowler95}{UKQCD~Collaboration, K.~C.~Bowler {\it et al.}}, Phys.~Rev.~D {\bf 51}, 4905 (1995)
\bibitem{ukqcd}{UKQCD~Collaboration, K.~C.~Bowler {\it et al.}}, Phys.~Rev.~D {\bf 62}, 054506 (2000)
\bibitem{stech}{M.~Wirbel, B.~Stech, M.~Bauer}, Z.~Phys.~C {\bf 29} 637 (1985)
\bibitem{brho95}{UKQCD~Collaboration, J.~M.~Flynn {\it et al.}}, Nucl.~Phys.~B {\bf 461} 327 (1996)
\bibitem{cleo}{CLEO~Collaboration, B.~H.~Behrens {\it et al.}}, Phys.~Rev.~D {\bf 61}, 052001 (2000)
\end{thebibliography}
\normalsize
\end{document}